\documentclass[12pt,preprint]{aastex}

\def\arcsecpoint{$''\!.$}

\received{}
\accepted{}

\slugcomment{To appear in the {\it The Astrophysical Journal}}

\shorttitle{NGC~3227}
\shortauthors{Crenshaw et al.}

\begin{document}

\title{Absorption and Reddening in the Seyfert Galaxy NGC 3227
\altaffilmark{1}}

\author{D.M. Crenshaw\altaffilmark{2},
S.B. Kraemer,
F.C. Bruhweiler,
\& J.R. Ruiz
}

\affil{Catholic University of America and Laboratory for Astronomy and
Solar Physics, NASA's Goddard Space Flight Center, Code 681,
Greenbelt, MD  20771}

\altaffiltext{1}{Based on observations made with the NASA/ESA Hubble Space 
Telescope. STScI is operated by the Association of Universities for Research in 
Astronomy, Inc. under NASA contract NAS5-26555. }

\altaffiltext{2}{crenshaw@buckeye.gsfc.nasa.gov}

\begin{abstract}

We have obtained low-dispersion spectra of NGC~3227 with the Space Telescope 
Imaging Spectrograph (STIS) to study the intrinsic UV absorption and the 
reddening of the nucleus in this Seyfert 1 galaxy. The UV spectra show a wealth 
of absorption lines at the systemic redshift that span a wide range in 
ionization state (Mg~I to N~V). The equivalent widths of the lines are 
consistent with our earlier prediction that a ``lukewarm absorber'' (T$_{e}$ $=$ 
18,000 K at the ionized face) with a 
substantial column of gas (N$_H$ $=$ 2 x 10$^{21}$ cm$^{-2}$) is present and 
likely responsible for the reddening of the nucleus. The lukewarm absorber is 
also responsible for most of the absorption in the X-rays at energies less than 
1 keV,  although a more highly ionized ``warm absorber'' is needed to account 
for the O~VII and O~VIII ionization edges. In addition, we require a small 
column (N$_H$ $=$ 5 x 10$^{19}$ cm$^{-2}$) of cold gas to match the strengths of 
the neutral and singly-ionized lines in the UV spectra. NGC 3227 is the first 
Seyfert galaxy in which a strong link between the reddening and intrinsic UV 
absorption has been found. By comparing our STIS UV and optical spectra  with 
those of the unreddened Seyfert NGC 4151, we have determined a reddening curve 
for the nuclear continuum source in NGC 3227 over the 1150 -- 10,200 \AA\ range. 
The reddening curve does not show a 2200 \AA\ bump, and is steeper in the UV 
than reddening curves derived for the Galaxy, LMC, and SMC, suggesting a 
preponderence of small dust grains near the nucleus.

\end{abstract}


~~~~~

\newpage

\section{Introduction}

The nearby Seyfert 1.5 galaxy NGC~3227 (z $=$ 0.00386) has been studied 
extensively in most regions of the electromagnetic spectrum. The UV and optical 
emission lines and continuum from the active nucleus are known to be reddened by 
dust in the Seyfert galaxy, although the derived values show a large dispersion 
(Komossa \& Fink 1997a). For example, Cohen (1983) determined E$_{B-V}$ $=$ 0.51 
from the narrow H$\alpha$/H$\beta$ ratio, whereas Winge et al. (1995) used the 
total (narrow $+$ broad) H$\alpha$/H$\beta$ ratio to obtain E$_{B-V}$ $=$ 0.28 
(the Galactic reddening is only E$_{B-V}$ $=$ 0.02,
Schlegel et al. 1998). {\it IUE} spectra 
show that the UV continuum is heavily reddened, as evidenced by the steep drop 
in continuum flux from the near to far UV (Courvoisier \& Paltani 1992).
X-ray observations of NGC~3227 with {\it ROSAT} and {\it ASCA} reveal the 
presence of an X-ray warm absorber, characterized by an ionization parameter 
(number of ionizing photons with energies $>$ 1 Ryd per hydrogen atom at the 
ionized face) of U $=$ 2.4 and hydrogen (H~I plus H~II) column density N$_{H}$ 
$=$ 3 x 10$^{21}$ cm$^{-2}$ (George et al. 1998). {\it IUE} observations {\it 
suggest} the presence of intrinsic UV absorption in Mg~II, although these 
observations cannot rule out a Galactic origin (Ulrich 1988). 

As pointed out by Komossa \& Fink (1997a), the intrinsic neutral column of gas 
in the line of sight to NGC~3227 (N$_{H I}$ $<$ 3 x 10$^{20}$ cm$^{-2}$, see 
also George et al. 1998) is not sufficient to produce the observed reddening, 
assuming a Galactic dust/gas ratio (N$_{HI}$ = 5.2 x 10$^{21}$ E$_{B-V}$ 
cm$^{-2}$, Shull \& van Steenberg 1985). Thus, Komossa \& Fink suggest that the 
reddening 
arises in the much higher column of highly ionized gas responsible for the X-ray 
absorption. Similar cases for ``dusty warm absorbers'' have been made for the 
QSO IRAS 13349$+$2438 (Brandt, Fabian, \& Pounds 1996) and the Seyfert galaxies 
NGC~3786 (Komossa \& Fink 1997b), IRAS 17020$+$4544 (Leighly et al. 1997), and 
MCG $-$6$-$30$-$15 (Reynolds et al. 1997).

In a recent paper on the reddening in X-ray absorbed Seyfert 1 Galaxies (Kraemer 
et al. 2000, hereafter K2000), we presented an alternative to the dusty warm 
absorber model. In our model, the dust exists in a ``lukewarm absorber'' in 
which the hydrogen is ionized, but the columns of O~VII and O~VIII are 
negligible. 
The lukewarm absorber can be placed at relatively large distances ($>$ 100 pc) 
from the nucleus, and therefore account for not only the reddening of the 
continuum source and broad-line region (BLR), but the reddening of the 
narrow-line region (NLR) as well.

In K2000, we showed a specific model for a lukewarm absorber in NGC 3227. To 
constrain the model, we required 1) the distance from the continuum source to be 
greater than $\sim$100 pc, to account for the reddening of the majority of the 
NLR emission, and 2) the column density to be N$_H$ $=$ 2 x 10$^{21}$ cm$^{-2}$, 
to account for the average reddening of  E$_{B-V}$ $=$ 0.4 from the literature 
(assuming the Galactic dust to gas ratio). The parameters of the lukewarm 
absorber were then determined from a fit to the soft X-ray absorption observed 
by {\it ASCA} and {\it ROSAT} (see K2000 for details). The resulting model was 
characterized 
by the parameters U $=$ 0.13, n$_H$ $=$ 20 cm$^{-3}$, and distance $=$ 120 pc. 
Interestingly, most of the soft X-ray absorption can then be attributed to 
He~II Lyman continuum absorption. Our lukewarm absorber model for NGC~3227 
predicted large ionic columns of N~V, C~IV, Si~IV, and Mg~II, which we suggested 
could be tested with observations in the UV at higher sensitivity than possible 
with {\it IUE}.

As part of an ongoing investigation into the intrinsic UV absorption in Seyfert 
galaxies, we obtained spectra of NGC~3227 with the Space Telescope Imaging 
Spectrograph (STIS) on the {\it Hubble Space Telescope} ({\it HST}) under a STIS 
guaranteed time observations (GTO) program. Given the {\it HST} time available 
for this object (2 orbits), the UV spectrum is too faint to obtain reasonable 
signal-to-noise levels with the echelle gratings, so we opted for low-dispersion 
UV spectra through a long slit (a description of STIS and its instrumental 
configurations can be found in Leitherer et al. 2000). We also obtained 
optical spectra through the same slit to investigate the reddening of the 
continuum over a broad wavelength region. 

\section{Observations and Measurements}

We observed the nucleus of NGC 3227 on 2000 February 8 using a 52$''$ x 
0\arcsecpoint2 slit and the STIS low-dispersion gratings. We show the details of 
the observations in Table 1. We reduced the spectra using the IDL software 
developed at NASA's Goddard Space Flight Center for the STIS Instrument 
Definition Team. The nucleus dominates the light in the long-slit images. We 
extracted the nuclear spectra using a 0\arcsecpoint3 height (perpendicular to 
the dispersion). We then combined the reduced spectra from each grating in their 
regions of overlap to produce a final spectrum over the 1150 -- 10,200 \AA\ 
range at a spectral resolving power of $\lambda$/$\Delta\lambda$ $\approx$ 1000.

Figure 1 shows the STIS UV/optical spectrum of the nucleus in NGC~3227 and that 
of the Seyfert 1.5 galaxy NGC 4151 (Nelson et al. 2000), scaled by a factor of 
0.15 in flux. The severe extinction in the UV spectrum of NGC 3227, compared 
to NGC 4151, is immediately obvious. The smooth curves in Figure 1 are spline 
fits to continuum regions that are free of any obvious contamination by 
absorption or emission lines including the ``little blue bump'' between 2200 
and 4200 \AA\ in the NGC 4151 spectrum, which consists of a blend of broad Fe~II 
and Balmer recombination emission (Wills, Netzer, \& Wills 1985). This feature 
may also be present in the NGC~3227 spectrum, but since it cannot be 
distinguished as a separate feature from the continuum emission, the continuum 
fit in this region may be considered an upper limit.

In Figure 2, we show an expanded view of the UV spectrum of NGC~3227. In 
addition to the broad and narrow emission lines and the interstellar absorption 
lines from the Galaxy, a number of intrinsic absorption lines are present and 
identified. These absorption lines appear at the systemic redshift of NGC~3227 
(z = 0.00386 from H~I 21 cm observations of the host galaxy, de Vaucouleurs et 
al. 1991), and span a wide range in the ionization potential needed to create 
these ions, from zero (O~I, C~I, Mg~I) to 77.5 eV (N~V). All of the absorption 
lines are unresolved at the velocity resolution of the spectra ($\sim$ 300 km 
s$^{-1}$ FWHM).

To measure the absorption lines, we fit cubic splines to adjacent wavelength 
regions, which typically contain both continuum and emission-line contributions.
Since the absorption lines are unresolved, the column densities cannot be 
determined directly from integration of optical depths (e.g., Crenshaw et al. 
1999), and we must therefore rely on the equivalent widths and 
radial velocity centroids (relative to the systemic redshift). A few of the 
absorption lines that we have identified are blended with stronger 
lines, and we are not able to determine reliable values for these lines. For the 
blended C~IV doublet, we were able to model the observed blend with two lines of 
approximately equal equivalent width. We show our absorption-line measurements 
in Table 2 (along with column densities from models discussed in section 4).

\section{The Reddening Curve for NGC~3227}

To determine a reddening curve from the nuclear continuum emission in NGC~3227, 
we assumed that 1) the intrinsic UV/optical continuum of NGC~4151 is essentially 
unreddened, and 2) the relative flux of NGC~4151 as a function of wavelength 
represents the intrinsic continuum spectrum of NGC~3227. The first assumption is 
based on the finding by Kriss et al. (1995) that the far-UV continuum of 
NGC~4151 is consistent with a power-law plus a small amount of reddening that is 
due solely to our Galaxy (E$_{B-V}$ $=$ 0.04, Burstein \& Heiles 1982).
The second assumption is based on evidence that the intrinsic continua of 
Seyfert 1 galaxies have a relatively standard shape, which is modified by the 
effects of dust and contributions from the host galaxy to produce the observed 
spectral energy distributions (Ward et al. 1987). We will explore the 
uncertainties associated with our assumptions later in this section.

Given our assumptions, we can determine the value of E(B$-$V) $=$ 2.5 
[log(X$_{B}$)~$-$~log(X$_{V}$)], where X is the ratio of the continuum fit to 
NGC~4151 to that of NGC~3227 as a function of wavelength (Figure 1), and is 
evaluated at the effective wavelengths of the B (4400 \AA) and V (5500 \AA) 
filters. Thus, we find that E(B$-$V) $=$ 0.18. We can also determine the 
reddening curve at any wavelength, relative to V and as per convention 
normalized to E(B$-$V) (Savage \& Mathis 1979):

\begin{equation}
\frac{E(\lambda - V)}{E(B - V)} \equiv
\frac{A_{\lambda} - A_{V}}{A_{B} - A_{V}} =
\frac{log(X_{\lambda}) - log(X_{V})}{log(X_{B}) - log(X_{V})},
\end{equation}

where A$_{B}$, A$_{V}$, and A$_{\lambda}$ are the extinctions in magnitudes at 
B, V, and an arbitrary $\lambda$.

Since we do not know the intrinsic continuum flux of NGC~3227 
at any wavelength, we cannot determine the total extinction directly. Thus, we 
need to calibrate the above reddening curve to determine the actual extinction 
law:
\begin{equation}
\frac{A_{\lambda}}{E(B-V)} = \frac{E(\lambda - V)}{E(B-V)} + R_{V},
\end{equation}

by determining the offset R$_{V}$ $\equiv$ A$_{V}$/E(B$-$V). The value of 
R$_{V}$ for the ``standard'' Galactic reddening curve is 3.1 (Savage \& Mathis 
1979), but this value can vary among different lines of sight in the Galaxy 
(Cardelli, Clayton, \& Mathis 1989). However, extinction laws appear to have 
essentially the same values at wavelengths $\lambda$ $>$ 7000 \AA\ 
(Cardelli et al. 1989). If we assume this is the case for NGC~3227, then R$_{V}$ 
is the constant we need to add to E($\lambda$$-$V)/E(B$-$V) to match the 
standard Galactic extinction law at long wavelengths.

Figure 3 shows extinction laws for the Galaxy, LMC, and SMC, along with the 
extinction law determined for NGC~3227 in the fashion described above. The shape 
of the extinction curve for NGC~3227 is very similar to the others over 
the 4000 -- 10000 \AA\ range, and we find that an offset of R$_{V}$ $=$ 3.2 
provides a good match to these curves; this value is very close to the standard 
Galactic value of R$_{V}$ $=$3.1.

The extinction curves in Figure 3 begin to differ 
strongly at $\lambda$ $<$ 4000 \AA. The standard Galactic curve (Savage \& 
Mathis 1979; Seaton 1979) shows a strong 2200 \AA\ bump, and the 
weakest (although substantial) rise to the UV. The LMC curve (Koornneef \& Code 
1981; Fitzpatrick 1985) shows a stronger rise to the UV and a less 
pronounced bump. The SMC curve shows the strongest rise to the UV (until now) 
and no detectable 2200 \AA\ bump (Hutchings et al. 1982; Witt \& Gordon 
1999).

The extinction curve for NGC~3227 is extreme, in that it shows a steeper rise to 
the UV than all of the other curves. Furthermore, the extinction curve resembles 
that of the SMC, in that there is no 2200 \AA\ bump. Interestingly,
Pitman, Clayton, \& Gordon (2000) find a similar result for higher luminosity 
AGN; specifically, there is no convincing evidence for a 2200 \AA\ feature in 
the spectrum of 
any QSO.
Tests show that lower continuum fits in the 2200 -- 4200 \AA\ region (assuming a 
substantial little blue bump, see section 2 and Figure 1) have little effect on 
the overall sharp rise to the far-UV. Given the reddening of E(B$-$V) $=$ 
0.18, we can determine the extinction of NGC~3227 at any wavelength from Figure 
1: e.g., A$_{\lambda}$(1200 \AA) $=$ 5.3, A$_{\lambda}$(2000 \AA) $=$ 2.9,
A$_{\lambda}$(3000 \AA) $=$ 1.2, and A$_{V}$(5500 \AA) $=$ 0.58. With this 
extinction curve, we find that the intrinsic continuum flux of NGC 3227 at any 
wavelength is then 0.17 times that of NGC~4151 at the respective times of 
observation. Note, however, that the STIS spectrum of NGC~3227 was obtained at a 
relatively low continuum state (section 4.2), whereas that of NGC~4151 was 
obtained at a moderately high state (Crenshaw et al. 2000).

To test the uniqueness of our reddening curve for NGC~3227, we have explored 
possible systematic uncertainties associated with our assumptions. The 
assumption that NGC 4151 represesents the intrinsic continuum spectrum of NGC 
3227 is based on the observed similarity of UV/optical continua in Seyfert 1 
galaxies (e.g., Wills, Netzer, \& Wills 1985). Unfortunately, {\it HST} 
observations of other unreddened Seyfert 1 galaxies covering the full UV/optical 
range at high spatial resolution (to avoid galaxy contamination) are not yet 
available. Therefore, we use the extreme variations of NGC~4151 itself, which 
was monitored extensively with {\it IUE}. Clavel et al. (1990) find that the 
ratio of UV to optical continuum F$_{\lambda}$(1450)/F$_{\lambda}$(5000) varies 
by a factor of 3.1, with high ratios corresponding to high UV fluxes. However, 
Kaspi et al. (1996) find that a portion of this effect is due to a substantial 
galaxy contribution in the apertures used for optical observations; subtracting 
the stellar flux at 5000 \AA\ leads to a ratio that varies by a factor of 2.1. 
Thus, if we assume an extremely high state for NGC 4151 and an extremely low 
state for NGC 3227, this factor would bring the reddening curve of NGC 3227 
close to (but still above that) of the SMC, since the reddening curves of these 
two objects in Figure 3 differ by 1.0 mag (on average) in the 1200 - 2000 \AA\ 
range (for E$_{B-V}$ $=$ 0.18). Other systematic uncertainties will tend to 
steepen the reddening curve slightly. Assuming that the entire 2500 -- 4000 bump 
in the observed spectrum (Figure 1) is due to the little blue bump leads to an 
artificial bump in the extinction curve that spans this region (with an 
amplitude of 0.3 mag), which does not resemble the narrower 2200 \AA\ bump in 
the standard Galactic curve. Scattering of the radiation back into our line of 
sight could potentially ``bluen'' the continuum, and correction for this effect 
would steepen the reddening curve in the UV. However this effect must be small, 
since our extraction slit covers only a 15 pc x 22 pc region, and we have 
estimated the dusty gas cloud is $\sim$100 pc from the nucleus. Assuming a 
spherical shell with a covering factor of one and uniform scattering, only 
$\sim$ 1\% of the UV light scattered in our direction enters the aperture. 
Furthermore, dust grain models show that most of the UV radiation is absorbed 
and re-emitted in the infrared, rather than scattered (Draine \& Lee 1984).
Thus, our estimate of the systematic uncertainties show that we cannot rule out 
a reddening curve for NGC 3227 that is essentially the same as that of the SMC, 
although it is likely to be steeper.

The STIS spectra also provide an opportunity to determine the reddening of the 
emission lines from the inner NLR (within 0\arcsecpoint15 or 11 pc of the 
nucleus for H$_{o}$ = 75 km s$^{-1}$ Mpc$^{-1}$) for comparison with the 
continuum reddening. At the temperatures and densities of the NLR, the He~II 
$\lambda$1640 and $\lambda$4686 lines are due to recombination, and the 
intrinsic He~II $\lambda$1640/$\lambda$4686 ratio is $\sim$7.2 (Seaton 1978; 
Kraemer et al. 1994).  We were able to deblend the narrow components of these 
lines from their broad counterparts, using the [O~III]$\lambda$5007 lines as a 
template, to determine an observed ratio of 0.19 $\pm$0.05. This yields 
A$_{\lambda}$(1640) $-$  A$_{\lambda}$(4686) $=$ 3.9 $\pm$0.3 mag for the narrow 
lines. By comparison, the continuum extinction curve yields A$_{\lambda}$(1640) 
$-$  A$_{\lambda}$(4686) $=$ 3.4. These values are in reasonably good agreement, 
and provide further evidence that the reddening of the continuum and emission 
lines is primarily due to material outside of the inner NLR (i.e., our lukewarm 
absorber). The small additional reddening experienced by the emission lines may 
be due to dust in the inner NLR, which has been detected in other Seyfert 
galaxies (e.g., NGC 1068, Kraemer \& Crenshaw 2000).

Our value for E(B$-$V) in NGC 3227 ($=$ 0.18) is somewhat lower 
than previous estimates (section 1). However, previous estimates 
have assumed the standard Galactic extinction curve, which is clearly incorrect, 
and the E(B$-$V) values are therefore sensitive to the wavelengths of the 
reddening indicators used. Also, the use of narrow H$\alpha$/H$\beta$ is 
complicated by the deconvolution of the narrow and broad components, and in 
particular, the deconvolution of narrow H$\alpha$ and [N~II] $\lambda\lambda$ 
6548, 6884, which may account for at least part of the large dispersion in
E(B$-$V) values. (Unfortunately, we cannot obtain a narrow H$\alpha$/H$\beta$ 
ratio from our own observations, because the spectral resolution of our own data 
does not allow an accurate deconvolution of the H$\alpha$ and [N~II] lines.) 
Finally, we note that previous reddening measurements were obtained for much 
larger projected regions ($>$~1$''$), and the reddening may be somewhat aperture 
dependent.

\section{The Intrinsic UV Absorption Lines}

\subsection{Properties of the Absorbers}

The high-ionization absorption lines in Figure 2 show significant saturation, 
suggesting a large column of gas. In particular, the C~IV 
$\lambda\lambda$1548.2, 1550.8 lines are blended and approach zero intensity in 
their troughs, 
despite the low resolution which decreases the observed depths of the cores of 
unresolved absorption lines. The N~V $\lambda\lambda$1238.8, 1242.8 and S~IV 
$\lambda\lambda$1393.8, 1402.8 absorption lines in Figure 2 also suggest a 
substantial column of ionized gas, in that the longer wavelength member of each 
doublet is much stronger than the 1:2 ratio for unsaturated lines. 
Qualitatively, these lines confirm our prediction in K2000 that a lukewarm 
absorber is present in NGC~3227.

Low-ionization absorption lines are relatively rare in Seyfert 1 galaxies; in an 
{\it HST} survey, we found that only one of 10 objects with 
intrinsic C~IV and N~V absorption showed Mg II in absorption (Crenshaw et al. 
1999). This object is NGC~4151, which shows not only low-ionization species, but 
absorption from fine-structure and metastable levels as well in most kinematic 
components (Kraemer et al. 2001, and references therein), indicating a range of 
moderate to high densities (n$_e$ from 10$^2$ cm$^{-3}$ to $>$10$^9$ cm$^{-3}$). 
For NGC~3227, the fine structure lines of O~I, C~II, Si~II, and Fe~II are not 
detected, indicating low densities (n$_e$ $<$ 100 cm$^{-3}$). Note that our 
original model in K2000 predicted negligible ionic columns of C~I and Fe~II. Our 
detection of these lines suggests the presence of a small column of cold gas, as 
discussed in subsequent sections.

The intrinsic absorption lines listed in Table 2 are near the systemic redshift 
of the host galaxy: the average radial velocity of all of the lines is $+$20 
$\pm$ 125 km s$^{-1}$. Although most intrinsic absorbers in Seyfert 1 galaxies 
are in outflow, systems near the systemic redshift are not uncommon (Crenshaw et 
al. 1999). There is some evidence that the high-ionization lines 
are at a slightly different radial velocity than the low-ionization lines: the 
average radial velocity for N~V, C~IV, and Si~IV is $-$70 $\pm$ 55 km s$^{-1}$, 
whereas the average radial velocity for the remaining, low-ionization lines is 
$+$125 $\pm$ 50 km s$^{-1}$.
Future observations at higher dispersions would be helpful in identifying 
multiple kinematic components and confirming this velocity difference.

\subsection{Photoionization Models of the Absorbers}

We have generated photoionization models to match the observed equivalent widths 
of the absorption lines.
The photoionization code has been described in previous publications
(cf. Kraemer et al. 1994). The photoionization models are parameterized in terms 
of the number of ionizing photons per hydrogen atom at the ionized face of the
absorber ($U$), the density (n$_{H}$), the distance from the ionizing source
(D), and the sum of the H~I and
H~II column densities (N$_{H}$). We assume that gas is ionized by
the continuum radiation emitted by the central source in the active
nucleus of NGC 3227. As discussed in K2000, we modeled the 
EUV to X-ray spectral energy distribution (SED) as a series of power-laws of the 
form F$_{\nu}$ $\propto$ $\nu^{-\alpha}$, with $\alpha = 1$ below 13.6~eV, 
$\alpha = 2$ over the range 13.6~eV $\leq h\nu <$ 500~eV, 
and $\alpha = 0.6$ above 500~eV. Note that the spectral
index above 500 eV matches the value derived from the best fit to the
combined {\it ROSAT}/PSPC and {\it ASCA} datasets (George et al. 1998). Based on 
this SED, 
and the assumption that NGC 3227 and NGC 4151 have similar optical-UV 
SEDs, we derived a luminosity in
ionizing photons, from 13.6 eV to 10,000 eV, of $\sim$ 1.5 x 10$^{53}$
photons s$^{-1}$. We assumed roughly solar element abundances (cf. Grevesse \& 
Anders 1989), which are, by number relative to H, as follows: He $=$ 0.1, 
C $=$ 3.4 x 10$^{-4}$, N $=$ 1.2 x 10$^{-4}$, O $=$ 6.8 x 10$^{-4}$,
Ne $=$ 1.1 x 10$^{-4}$, Mg $=$ 3.3 x 10$^{-5}$, Si $=$ 3.1 x 10$^{-5}$,
S $=$ 1.5 x 10$^{-5}$, and Fe $=$ 4.0 x 10$^{-5}$. Further,
we assumed that both silicate and carbon dust grains are present in the gas, 
with a power-law distribution in sizes (see Mathis, Rumpl, \& Nordsieck 1977).
For our original lukewarm model, we assumed depletions of elements from gas 
phase typical of the Galactic interstellar medium (cf. Snow \& Witt 1996): 
C, 65\%; O, 50\%; Si 95\%; Mg 82\%; and Fe 95\% (K2000). 

In K2000, we determined that the observed reddening
and the bulk of the X-ray absorption could be explained by a single
component of ionized gas (a lukewarm absorber), which covers the
continuum source, broad-line region, and much of the narrow-line region.
As discussed in Section 1 and K2000, we had previously fixed the total column 
density of the lukewarm absorber at N$_{H}$ $=$ 2 x 10$^{21}$ cm$^{-2}$,
based on the reddening derived from the Balmer decrement and the Galactic
dust-to-gas ratio. From our empirical reddening curve for NGC 3227, we have
determined E(B$-$V) $=$ 0.18, roughly half the value we assumed
in K2000. and consequently one might argue that N$_{H}$ should be reduced by 
the same factor. However, since the 
reddening curve is quite different in NGC 3227 than in the Galaxy, it
is unlikely that the Galactic relationship of the reddening to the total 
hydrogen column strictly applies. Furthermore, our success in fitting the 
soft X-ray absorption gives us confidence that the He~II column density,
and hence the U and N$_{H}$ are roughly correct. Therefore, we have
opted to keep the same parameters as before for the lukewarm component:
$U$ $=$ 0.13, n$_{H}$ $=$ 20 cm$^{-3}$, and N$_{H}$ $=$ 2 x 10$^{21}$ cm$^{-2}$.
Note that the distance of the lukewarm absorber from the central source depends 
on the luminosity, and we derived D $=$ 120 pc from our original finding that 
the reddening-corrected luminosity of NGC~3227 derived from the 1992 spectra in 
Winge et al. (1995) is equal to that of NGC~4151 from Nelson et al. (2000). 
However, the smaller reddening derived in this paper places the 
luminosity of NGC 3227 at 0.59 times that of NGC 4151, which yields D $=$  
92 pc. Furthermore, we have assumed that the 1992 luminosity represents the 
time-averaged value, although the current STIS spectrum indicates that it is in 
a lower continuum state than in previous observations (i.e., the continuum flux 
at 5500 \AA\ is 30\% that of the average value from the 1992 Winge et al. 
spectra).

As noted in section 4.1, the STIS spectra reveal the presence of 
saturated N~V and C~IV lines, in qualitative agreement
with the predictions of the lukewarm model. However, the equivalent widths of 
the Si~IV $\lambda\lambda$1393, 1403 lines are about twice those 
expected from our original model and a curve-of-growth that fits the 
C~IV and N~V lines (see the next subsection), which is evidence for a greater 
column of 
Si~IV than we had predicted. We suggest that this is the result of a smaller 
fractional depletion of silicon onto dust grains. 
This is not surprising, given the unique extinction law we have determined for
NGC 3227. However, since we do not have an independent way to determine the
elemental depletions for the constituents of the silicate grains, we have 
recomputed the lukewarm model using the Galactic depletions, scaled by 0.8 to 
account for the larger fraction of silicon in gas-phase, as follows: O, 41\%; 
Si, 77\%; Mg, 67\%;  and Fe, 78\%. While the 2200 \AA~ bump is generally assumed 
to be due to carbon-based grains, the origin of this feature is not
well understood (see Duley \& Seahra 1999, and references therein) and
its relative strength may not be correlated with the amount of carbon depletion 
(Cardelli et al. 1996).
Hence, we have assumed the same depletion of carbon as in K2000. 

As we will demonstrate in section 4.3,
the lukewarm model predicts relatively small ionic columns for most
low-ionization species (and essentially no Fe~II), although such lines
are strong in our STIS spectra (Table 2). We attempted to
increase the low ionization column densities by increasing N$_{H}$,
but models that fit the observed EW's of the Si~II, Mg~II, and Fe~II lines
underpredict those of the O~I and C~II lines. Furthermore, there is some
evidence that the low ionization lines have a different velocity radial
centroid than that of the high ionization lines (section 4.1). Hence, we 
have included an additional component of low ionization gas (a ``cold 
component'').
Based on the absence of fine structure lines of Si~II and C~II, the density of 
the cold component is $<$~100 cm$^{-2}$ and, consequently, cannot be
closer to the source of ionizing radiation than the lukewarm gas 
(for example, if the two components were at the same radial distance,
the cold component would still have a fairly high ionization parameter of U $=$ 
0.03). 
Therefore, we have placed the cold component further from the nucleus and, as a 
result, it is 
screened by the lukewarm gas. In modeling this component, we have assumed the 
same abundances and depletions
as the lukewarm component, and the following parameters: $U$ $=$ 0.00015,
n$_{H}$ $=$ 100 cm$^{-2}$, N$_{H}$ $=$ 5.2 x 10$^{19}$ cm$^{-2}$, and D
$=$ 160 pc.
The column density was set to provide the best fit to the observed low 
ionization lines, as described in the next section.
The resulting ionic column densities for both lukewarm and cold components are 
given in Table 2.

\subsection{Comparison of Observations and Models}

To demonstrate that the equivalent widths of the absorption lines are consistent 
with the predictions of our absorber models, we have generated Maxwellian curves 
of growth for different velocity widths (``b'' values in km s$^{-1}$, see 
Spitzer et al. 1978), as shown in Figure 4. In this figure we overplot the 
absorption-line data from Table 2 (in a few cases, lines from the same ion have 
nearly identical values and only one line is plotted for clarity).
For each absorption line that is overplotted, the vertical position is 
determined by the observed equivalent width and the horizontal position is 
determined by the ionic column predicted by the model. For a group of lines to 
be consistent with a model, the lines should lie relatively near a single curve 
of growth representing a particular b value.

The top panel in Figure 4 shows our results for the lukewarm absorber model 
described in the previous section. This plot shows that the high-ionization 
lines (C~IV, N~V, and Si~IV) are consistent with the ionic column densities from 
the lukewarm model and a b value of 100 $\pm$ 50 km s$^{-1}$. Mg II also matches 
the curve of growth at b $=$ 100 km s$^{-1}$. However, the column densities of 
the low-ionization lines of C~II, Si~II and Mg~I are underpredicted, and the 
lukewarm model predicts negligible columns of O~I, C~I, and Fe~II. Thus, an 
additional component of relatively cold gas is required to match these lines.

The bottom panel in Figure 4 shows our result for the cold absorber model 
described in the previous section. Nearly all of the low-ionization lines are 
consistent with the ionic columns predicted by the model and
b $=$ 200 $\pm$ 50 km s$^{-1}$. However, the columns of Mg~I and Mg~II are 
overpredicted, 
which is exacerbated by the fact that some of the Mg~II equivalent width must be 
attributed to the lukewarm absorber. This suggests that the depletion of Mg into 
dust may actually be somewhat larger than the value chosen for our models (i.e., 
85\% rather than the 67\% we assumed), and 
that, in general, the relative depletion of the constituents of the dust 
grains differs from that in the Galactic ISM. The only 
low-ionization line that is severely underpredicted by this model is C~I.
Our underprediction of the C~I column is likely the result of the 
unusually steep extinction curve in the far UV, which is not accounted for in 
our model of the lukewarm absorber. The large 
extinction inferred at $\sim$1100 \AA\ depletes photons 
with energies near the ionization potential of C~I (11.26 eV), which would 
likely
result in a substantial C~I column within the slab (note only $\sim$1\% of the C 
needs to be C~I.)

Overall, the curves of growth in Figure 4 show that the equivalent widths of the 
absorption lines are consistent with our models 
of the lukewarm and cold absorbers in the line of sight to the nucleus. The 
different b values and possibly different velocity centroids for the low and 
high-ionization lines are further evidence for two components with different 
kinematics. Spectral observations of the absorption at higher dispersion, 
although challenging due to the low UV flux, would be helpful in providing 
further tests of and/or refinements to these models.

\section{Conclusions}

We have shown that there are three very different intrinsic components of gas in 
the line of sight to the nucleus of NGC 3227: an X-ray warm absorber (U $=$ 2.4, 
N$_H$ = 3 x 10$^{21}$ cm$^{-2}$), a lukewarm absorber (U $=$ 0.13, N$_H$ =2 x 
10$^{21}$ cm$^{-2}$), and a cold absorber (U $=$ 1.5 x 10$^{-4}$, N$_H$ = 5 x 
10$^{19}$ cm$^{-2}$. The X-ray warm absorber, characterized by O~VII and O~VIII 
absorption edges, shows a variable column of highly ionized gas, as discussed in 
detail by George et al. (1998).  The lukewarm absorber predicted by K2000 and 
revealed by the STIS spectra shows high-ionization lines (N~V, C~IV, Si~IV) 
typical of intrinsic UV absorbers in most Seyfert 1 galaxies (Crenshaw et al. 
1999). The low ionization lines in the STIS spectra are primarily due to the 
cold component, although our model indicates significant contributions to the 
observed C~II, Si~II, and Mg~II from the lukewarm absorber. The 
H~I column from the cold component is only 3 x 10$^{19}$ cm$^{-2}$ (Table 2), 
which is much smaller than the intrinsic neutral column of $\sim$3 x 10$^{20}$ 
cm$^{-2}$ derived from the soft X-ray absorption (Komossa \& Fink 1997a; George 
et al. 1998). However, as we pointed out in K2000, most of the soft X-ray 
absorption can be attributed to He~II Lyman continuum absorption from the 
lukewarm absorber, and thus previous estimates of the neutral column from X-ray 
studies of NGC~3227 are likely a factor of ten too large.

We find that our lukewarm absorber has sufficient column to provide the dust 
required to redden the continuum and emission lines in NGC~3227, assuming a 
Galactic dust-to-gas ratio. 
Furthermore, the lukewarm absorber is at a distance sufficient to account for 
the reddening of the NLR. Thus, we prefer the lukewarm absorber as the source of 
the reddening. Although a dusty warm absorber cannot be completely ruled out, 
the variability of the warm absorber in NGC~3227 over a time scale of two years 
suggests a location that is within light days of the central source (George et 
al. 1998), which is unsuitable for reddening the NLR.
Further monitoring would be helpful in verifying the variability and 
establishing the location of the X-ray absorber.

We note that additional support for reddening by dust at large distances comes 
from spectropolarimetry of NGC 3227. Schmidt \& Miller (1985) find that the 
continuum polarization increases from 1\% to 3\% going from 6000 \AA\ to 3500 
\AA. Furthermore, narrow emission lines are polarized by the same or slightly 
smaller amounts as nearby continuum regions, suggesting scattering by aligned 
dust grains at the location or outside of the NLR. This confirms the suggestion 
by Thompson et al. (1980) that the nucleus, BLR, and NLR of NGC~3227 are covered 
by a polarizing dust cloud.

The extinction curve that we have derived for NGC~3227 has important 
implications for reddening corrections to Seyfert galaxies. For example, 
photoionization models of the NLR rely heavily on UV/optical lines ratios (e.g., 
Ly$\alpha$/H$\beta$, C~IV/H$\beta$) to determine the physical conditions in the 
emitting gas (cf., Kraemer \& Crenshaw 2000). However, the determination of 
intrinsic 
line ratios is very sensitive to the reddening curve, which is almost always 
assumed to be Galactic. For E$_{B-V}$ $=$ 0.18 and the extinction curve for NGC 
3227 in Figure 3, the application of the standard Galactic curve would result in 
an underestimate of the extinction at 1200\AA\ of 3.6 mag (i.e., the corrected 
flux would be too small be a factor of 27.5). Note that 
this effect is mitigated if the He~II $\lambda$1640 and He~II $\lambda$4686 
lines are 
used to determine the reddening; the long UV/optical baseline provides a crude 
first-order correction, even if the wrong reddening curve is assumed. However, 
the use of the Galactic curve would still result in substantial errors, and in 
particular, an overestimate of the extinction in the 2200 \AA\ bump region.

The extinction curve also has important implications for the properties of the 
dust near the nucleus, since the shapes of the reddening curves depend on the 
composition and size distribution of the dust grains. Theoretical work in this 
area (Mathis, Rumpl, \& Nordsieck 1977; Draine \& Lee 1984) attributes the ratio 
of UV to optical absorption to the sizes of the grains (large UV absorption 
corresponds to smaller average grain size). Thus, the large UV extinction 
suggests a smaller average size of dust 
grains (compared to the Galaxy) for the SMC (Pei 1992; Weingartner \& Draine 
2000) and, by analogy, NGC 3227. One possible explanation for the smaller grain 
sizes near the nucleus of this Seyfert galaxy is that they are due to shocks, 
which tend to destroy large grains via shattering (Jones et al. 1996).

It is interesting that NGC 4151 harbors a UV absorber with 
a large column but shows no evidence for dust in the gas. STIS echelle 
observations of NGC~4151 show a kinematic component of absorption at $-$500 km 
s$^{-1}$ (relative to systemic) that is characterized by U $=$ 0.015 -- 0.060 
(variable) and N$_{H}$ $=$ 2.8 x 10$^{21}$ cm$^{-2}$ (Kraemer et al. 2001), but, 
as we have discussed, the UV/optical continuum of NGC 4151 shows no significant 
reddening. However, the high-column absorber in NGC 4151 is at a distance of 
only $\sim$ 0.03 pc from the nucleus (Kraemer et al. 2001), which is the 
approximate dust sublimation radius for the UV luminosity of this Seyfert 
(Barvainis 1987). Furthermore, given the large difference in distance from the 
nuclear continuum source, it is possible that the high-column UV absorbers in 
these two Seyferts have completely different origins.

{\it HST} observations of the (faint) UV emission from other reddened AGN would 
be helpful in exploring the relationship between absorption and reddening in 
these objects. Given that a substantial fraction of Seyfert 1 galaxies show 
evidence for reddening of their continua by dust (Ward et al. 1997), it would be 
interesting to determine their reddening curves with the procedures we 
have presented in this paper. New observations would be particularly useful for 
determining the range of intrinsic spectrum shapes in unreddened AGN and 
exploring the variety of extinction laws in reddened AGN.

\acknowledgments
We thank U.J. Sofia for helpful discusions on extinction laws and depletions and 
L. Magnani for useful discussions on high-latitude galactic clouds. We thank the 
anonymous referee for helpful suggestions.
This work was supported by NASA Guaranteed Time Observer funding to the STIS 
Science Team under NASA grant NAG 5-4103.

\clearpage

\figcaption[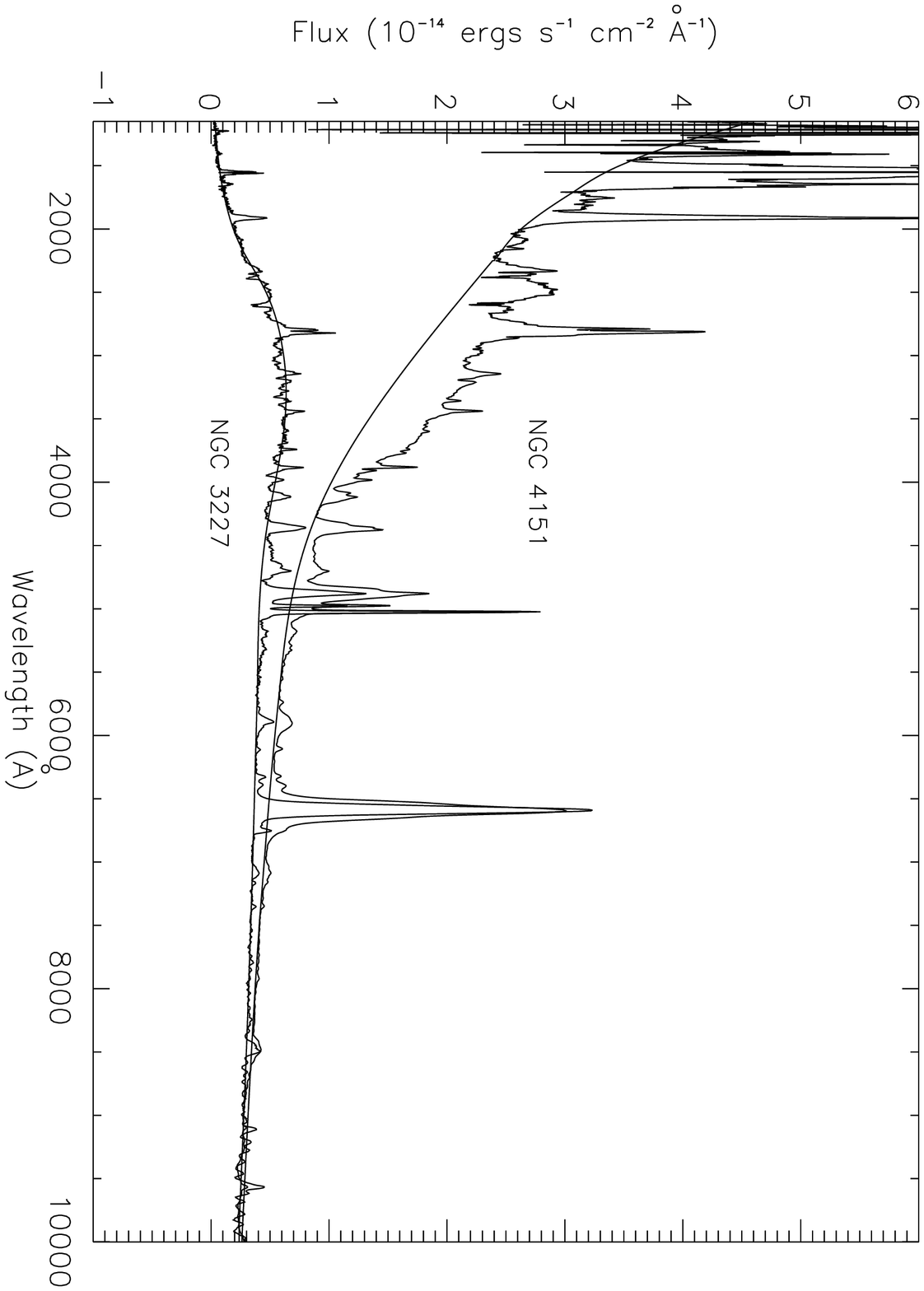]{STIS UV and optical spectra of NGC 3227 and NGC 4151 
(scaled in flux by a factor of 0.15). The smooth lines are spline fits to 
continuum regions in each spectrum.}

\figcaption[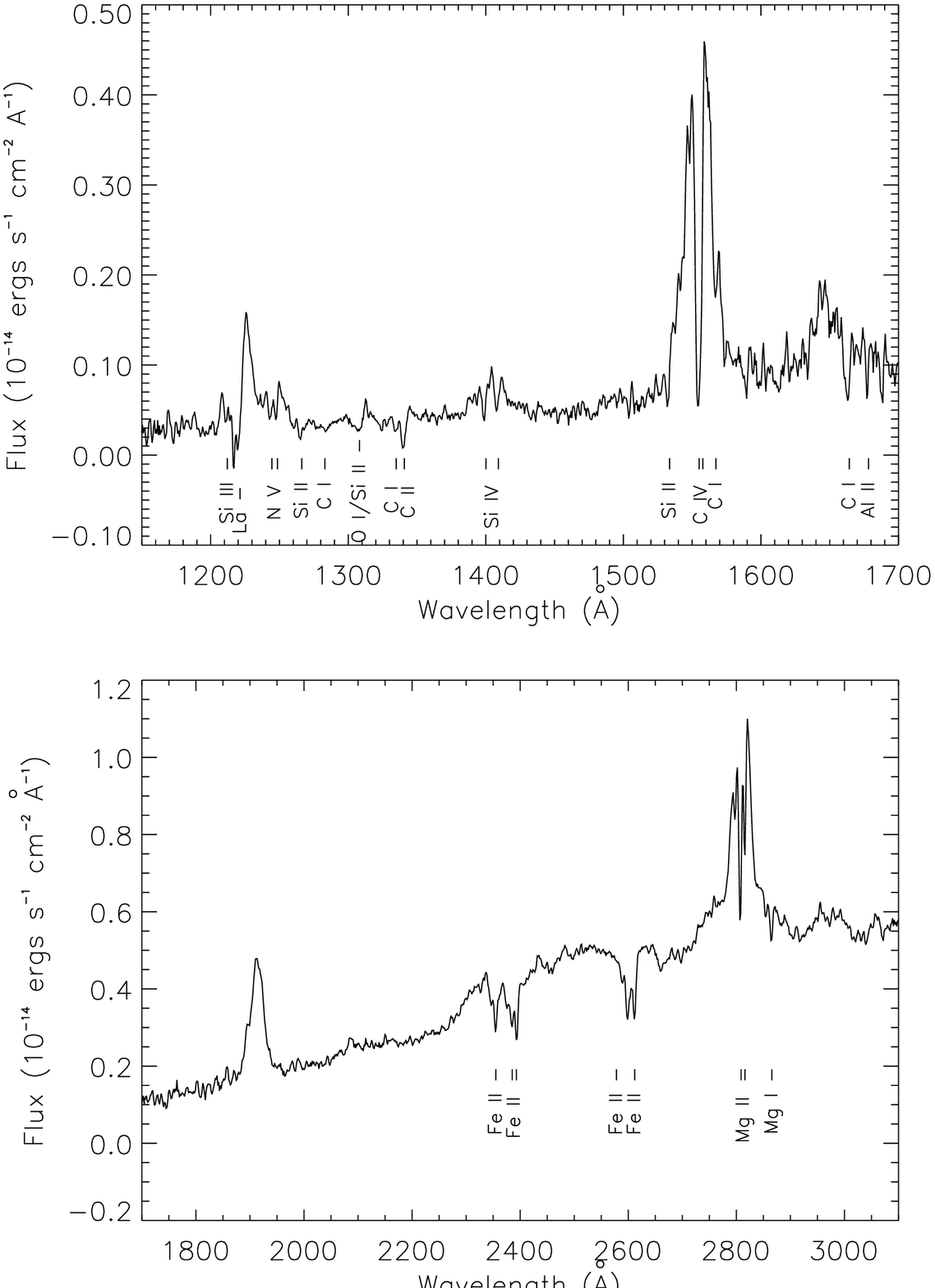]{Expanded view of the UV spectra of NGC 3227. Intrinsic 
absorption lines detected near the systemic redshift (z $=$ 0.00386) are 
labeled.}

\figcaption[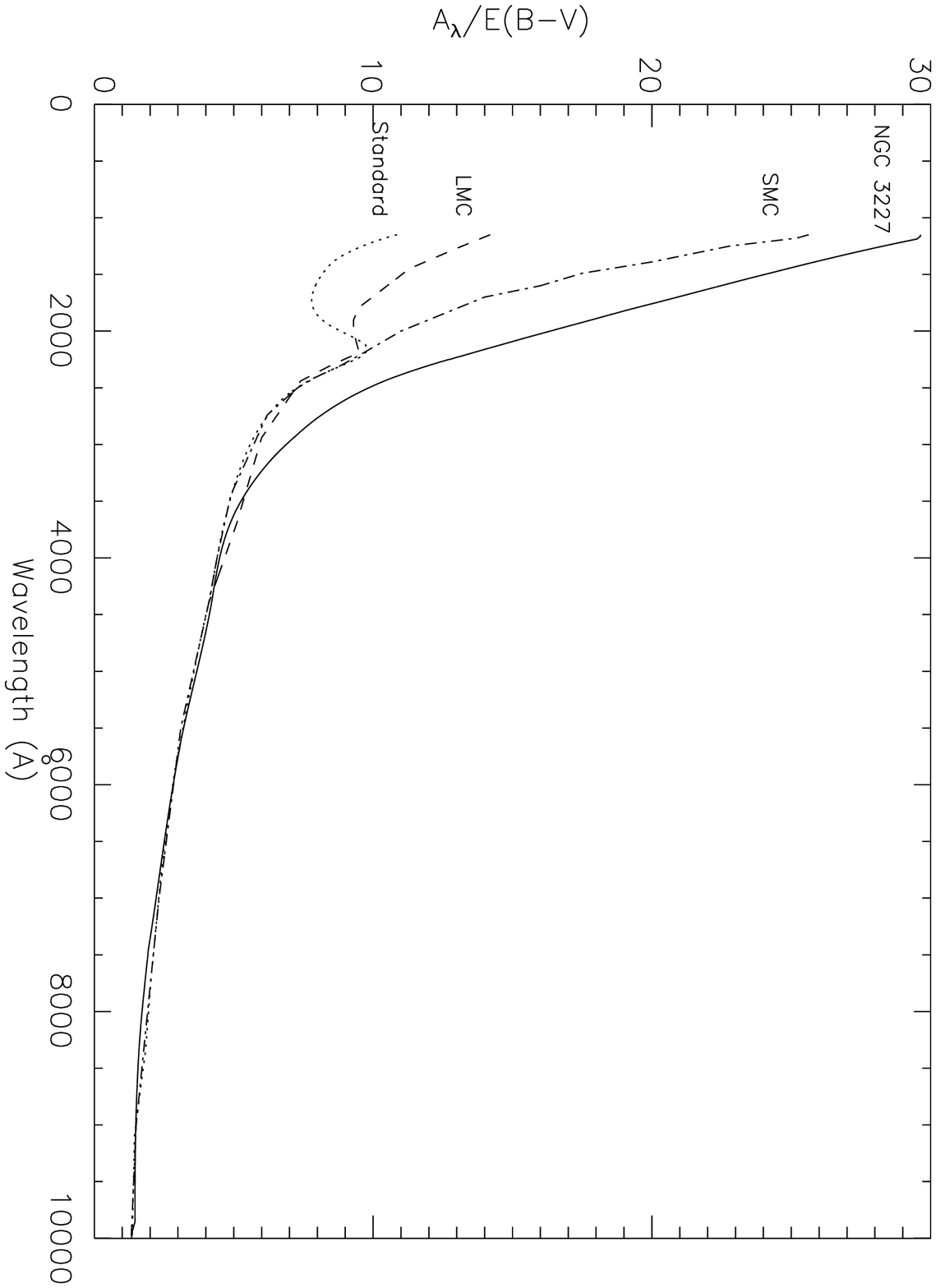]{Reddening curve determined for NGC 3227 as a function of 
wavelength. Reddening curves for the Galaxy (the standard curve of Savage \& 
Mathis 1979), LMC (Koornneef \& Code 1981) and SMC (Hutchings 1982) are given 
for comparison.}

\figcaption[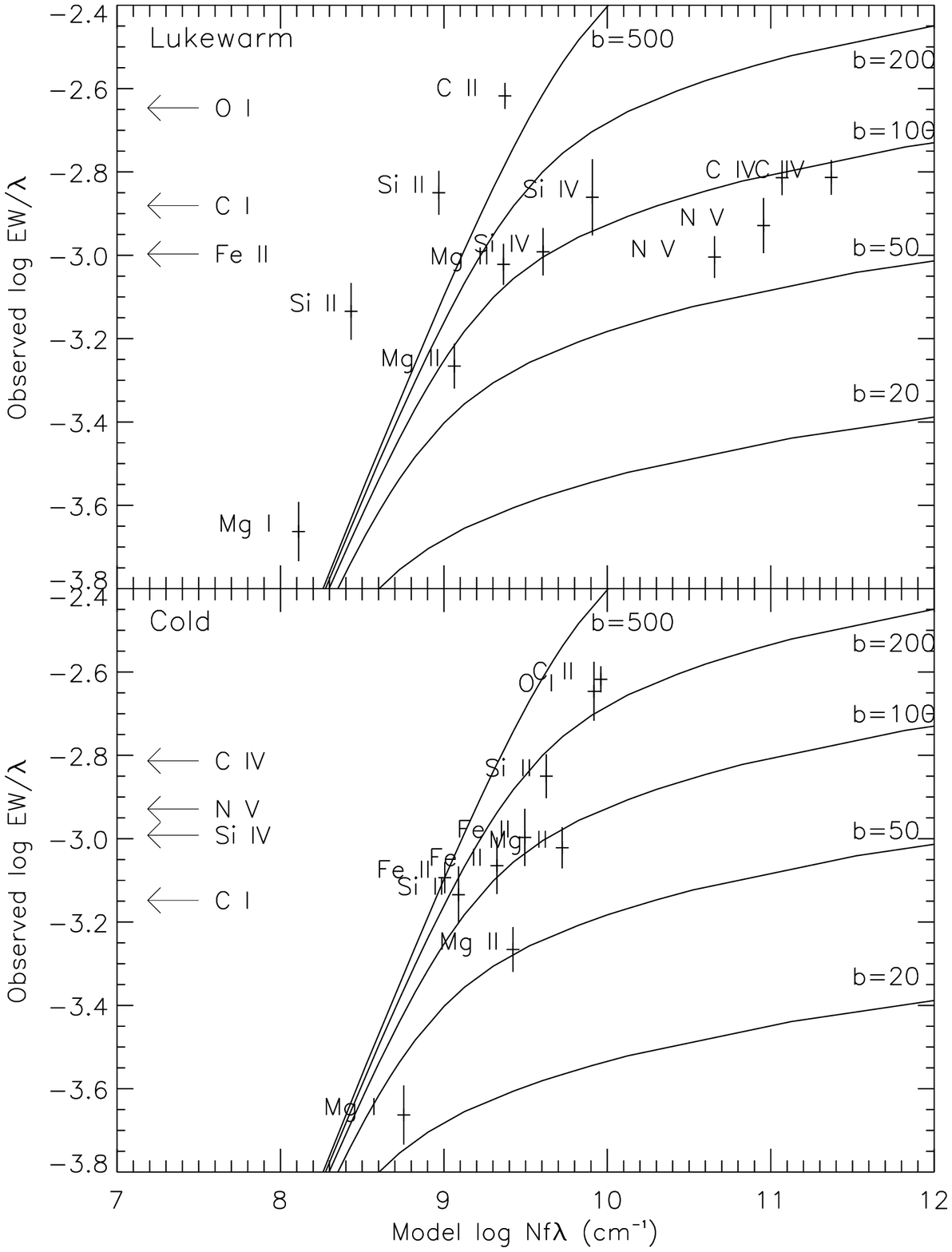]{Empirical curves of growth, based on a Maxwellian velocity 
distribution, for different velocity widths (``b'' values given in km s$^{-1}$, 
Spitzer et al. 1978). 
The vertical position of each point is determined from the observed equivalent 
width of an absorption line. The horizontal direction is determined by the 
column density for the corresponding ion from the model. The upper and lower 
panels give values based on our models of the lukewarm and cold absorption 
components, respectively.} 

\clearpage
\begin{deluxetable}{lcc}
\tablecolumns{3}
\footnotesize
\tablecaption{STIS Observations of NGC~3227}
\tablewidth{0pt}
\tablehead{
\colhead{Grating} & \colhead{Exp. Time} & \colhead{Coverage} \\
\colhead{} &\colhead{(sec)} &\colhead{(\AA)}
}
\startdata
G140L & 2132 & 1150 -- 1715\\
G230L & 1612 & 1573 -- 3157\\
G430L & 120  & 2900 -- 5710\\
G750L & 120  & ~5267 -- 10,261\\

\enddata
\end{deluxetable}

\clearpage
\begin{deluxetable}{lrcccl}
\tablecolumns{6}
\footnotesize
\tablecaption{Intrinsic Absorption -- Measurements and Models}
\tablewidth{0pt}
\tablehead{
\colhead{Line} & \colhead{v$_r$$^{a}$} & \colhead{EW$^{b}$} &
\colhead{N (Lukewarm)$^{c,d}$} & \colhead{N (Cold)$^{c,e}$} &
\colhead{Comments on}\\
\colhead{} &\colhead{(km s$^{-1}$)} &\colhead{(\AA)}
& \colhead{(cm$^{-2}$)} & \colhead{(cm$^{-2}$)} &
\colhead{Observations} 
}
\startdata
Si~III     $\lambda$1206.5 &        &                 &6.8 x 10$^{14}$
 &9.4 x 10$^{12}$ &blended with Ly$\alpha$\\
Ly$\alpha$ $\lambda$1215.7 &        &                 &1.8 x 10$^{18}$
 &3.0 x 10$^{19}$ &saturated\\
N~V   $\lambda$1238.8 & $-$44  &1.46 $\pm$0.24 &4.8 x 10$^{16}$
 &------------ & \\
N~V   $\lambda$1242.8 & $-$71  &1.23 $\pm$0.15 &4.8 x 10$^{16}$
 &------------ & \\
Si~II $\lambda$1260.4 & $+$103 &1.78 $\pm$0.23 &7.7 x 10$^{13}$
 &3.5 x 10$^{14}$ & \\
C~I   $\lambda$1277.2 & $+$145 &1.07 $\pm$0.39 &------------
 &3.4 x 10$^{13}$ & \\
O~I   $\lambda$1302.2 & $-$236 &2.94 $\pm$0.52 &1.1 x 10$^{13}$
 &1.3 x 10$^{16}$ & includes Si~II $\lambda$1304.4\\
Si~II $\lambda$1304.4 &        &               &7.7 x 10$^{13}$
 &3.5 x 10$^{14}$    & blended with O~I\\
C~I   $\lambda$1328.8 & $+$174 &0.99 $\pm$0.23 &------------
 &3.4 x 10$^{13}$ & includes Gal. C~II\\
C~II  $\lambda$1334.5 & $+$100 &3.22 $\pm$0.24 &1.5 x 10$^{15}$
 &5.8 x 10$^{15}$ & \\
Si~IV $\lambda$1393.8 & $-$38  &1.92 $\pm$0.45 &1.1 x 10$^{15}$
 &------------ & \\
Si~IV $\lambda$1402.8 & $-$46  &1.43 $\pm$0.20 &1.1 x 10$^{15}$
 &------------ & \\
Si~II $\lambda$1526.7 & $-$77  &1.12 $\pm$0.19 &7.7 x 10$^{13}$
 &3.5 x 10$^{14}$ & \\
C~IV  $\lambda$1548.2 & $-$170 &2.38 $\pm$0.24 &7.8 x 10$^{16}$
 &------------ & EW from fit to doublet\\
C~IV  $\lambda$1550.8 & $-$170 &2.38 $\pm$0.24 &7.8 x 10$^{16}$
 &------------ & EW from fit to doublet\\
C~I   $\lambda$1560.3 & $+$72  &1.11 $\pm$0.35 &------------
 &3.4 x 10$^{13}$ & \\
C~I   $\lambda$1656.9 & $-$162 &2.18 $\pm$0.34 &------------
 &3.4 x 10$^{13}$ & \\
Al~II $\lambda$1670.8 & $-$9   &0.85 $\pm$0.23 &               
 &                & \\
Fe~II $\lambda$2344.2 & $+$198 &1.89 $\pm$0.27 &------------
 &4.0 x 10$^{14}$ & \\
Fe~II $\lambda$2374.4 &        &               &------------
 &4.0 x 10$^{14}$ &blended with Gal. Fe~II\\
Fe~II $\lambda$2382.8 & $+$151 &2.40 $\pm$0.41 &------------
 &4.0 x 10$^{14}$ & \\
Fe~II $\lambda$2586.6 &        &               &------------
 &4.0 x 10$^{14}$ &blended with Gal. Fe~II\\
Fe~II $\lambda$2600.2 & $+$122 &2.24 $\pm$0.38 &------------
 &4.0 x 10$^{14}$ & \\
Mg~II $\lambda$2796.3 & $+$11  &2.66 $\pm$0.32 &1.4 x 10$^{14}$
 &3.2 x 10$^{14}$ & \\
Mg~II $\lambda$2803.5 & $+$115 &1.52 $\pm$0.20 &1.4 x 10$^{14}$
 &3.2 x 10$^{14}$ & \\
Mg~I  $\lambda$2853.0 & $+$61  &0.62 $\pm$0.11 &2.5 x 10$^{12}$
 &1.1 x 10$^{13}$ & \\

\tablenotetext{a}{relative to the systemic redshift of 0.00386}
\tablenotetext{b}{blank entry: blended with another line.}
\tablenotetext{c}{blank entry: not predicted by model. --- entry: negligible 
ionic column} 
\tablenotetext{d}{U $=$ 0.13, N(H I + H II) $=$ 2 x 10$^{21}$ cm$^{-2}$.}
\tablenotetext{e}{U $=$ 1.5 x 10$^{-4}$, N(H I + H II) $=$ 5.2 x 10$^{19}$ 
cm$^{-2}$.}

\enddata
\end{deluxetable}

%\end{document}

\clearpage
\vskip3.0in
\begin{figure}
\plotone{f1.eps}
\\Fig.~1.
\end{figure}

\clearpage
\vskip3.0in
\begin{figure}
\plotone{f2.eps}
\\Fig.~2.
\end{figure}

\clearpage
\vskip3.0in
\begin{figure}
\plotone{f3.eps}
\\Fig.~3.
\end{figure}

\clearpage
\vskip3.0in
\begin{figure}
\plotone{f4.eps}
\\Fig.~4.
\end{figure}
\end{document}